\documentclass[epsfig, usenatbib]{mn2e}
\usepackage{epsfig}
\usepackage{natbib}
\usepackage{graphicx}
\usepackage{fixltx2e}
\usepackage{multirow}
\usepackage{amsmath}
\usepackage{amssymb}
\usepackage{url}
\title{Disentangling star formation and merger growth in the evolution of Luminous Red Galaxies}

\author[Tojeiro et al.]{Rita Tojeiro\thanks{E-mail: rita.tojeiro@port.ac.uk}$^1$,  
  Will J. Percival$^1$\\
$^1$Institute of Cosmology and Gravitation, Dennis Sciama Building, University of Portsmouth,
Burnaby Road, Portsmouth, PO1 3FX \\
}

\def\gs{\mathrel{\raise1.16pt\hbox{$>$}\kern-7.0pt %
\lower3.06pt\hbox{{$\scriptstyle \sim$}}}}         %
\def\ls{\mathrel{\raise1.16pt\hbox{$<$}\kern-7.0pt %
\lower3.06pt\hbox{{$\scriptstyle \sim$}}}}         %

\date{Submitted to MNRAS}

\begin{document}

\maketitle
\begin{abstract}
  We introduce a novel technique for empirically understanding galaxy evolution. We use empirically determined stellar evolution models to predict the past evolution of the Sloan
  Digital Sky Survey (SDSS-II) Luminous Red Galaxy (LRG) sample
  without any a-priori assumption about galaxy evolution. By carefully contrasting the evolution of the predicted and
  observed number and luminosity densities we test the passive
  evolution scenario for galaxies of different luminosity, and
  determine minimum merger rates. We find that the LRG population is not purely coeval,
  with some of galaxies targeted at $z<0.23$ and at $z>0.34$ showing
  different dynamical growth than galaxies targeted throughout the
  sample. Our results show that the LRG
  population is dynamically growing, and that this growth must be
  dominated by the faint end. For the most luminous galaxies, we find
  lower minimum merger rates than required by previous studies that
  assume passive stellar evolution, suggesting that some of the
  dynamical evolution measured previously was actually due to galaxies
  with non-passive stellar evolution being incorrectly modelled. Our methodology can be used to identify and match
 coeval populations of galaxies across cosmic times, over one or more surveys.

\end{abstract}


\section{Introduction}  \label{sec:intro}

The most recent observational evidence points towards a model of structure and galaxy formation that is hierarchical in nature: small fluctuations in the matter density field grow via gravity, with their dynamics being governed in detail by dark matter and dark energy. Baryons trace the dark matter and, in regions of sufficient gravitational depth, they accumulate and form stars and galaxies \citep{WhiteRees78}. As the matter field continues to evolve dynamically - largely oblivious to this process -  the stellar content of galaxies grows via a combination of two modes: forming new stars from cold gas, and merging with other galaxies. 

LRGs are extensively used as cosmological probes (e.g. \citealt{ReidEtAl10, PercivalEtAl10}), and are also interesting from a galaxy evolution perspective, as  they dominate the galaxy mass function at the massive end. For these reasons, the stellar and dynamical evolution of LRGs and early-type galaxies (ETGs) has been studied extensively (see \citealt{TojeiroEtAl10, TojeiroEtAl11} - henceforth T11 - for a summary and list of references). Studies that address one of these growing modes, however, traditionally assume a model for the other. In this paper we propose and apply a new methodology that solves for these two modes independently. In brief, we use the fossil record of local galaxies to predict number and luminosity densities at past redshifts. The differences between the predicted and the observed quantities, under some simple assumptions, can be interpreted as a merger history.

The idea of inferring the properties of galaxies at a different cosmic time from the one they are observed at, and comparing them to the in-situ properties of galaxies at those cosmic times has been proposed before. Most notably \cite{DroryAlvarez08} use the galaxy stellar mass function (GSMF) of galaxies in the FORS Deep Field \citep{DroryEtAl05} and estimates of the instantaneous star-formation rate as a function of observed stellar mass to separate the evolution of the GSMF in terms of its merging and star formation components. Our approach differs from theirs in terms of the observables (we focus on number and luminosity densities, as opposed to the GSMF), but mainly in how we effectively link the galaxies at different cosmic times. We rely on the fossil record of local galaxies to reconstruct their past stellar build-up, which we get from VESPA \citep{TojeiroEtAl07} in a  non-parametric way. \cite{DroryAlvarez08} use parametric functions for the evolution of the GSMF and for the instantaneous star-formation rate, which they integrate over past cosmic time to predict a mass build up due to star-formation only. In other words, we use present-day information to predict the past Universe, whilst they use past information to predict the present-day Universe. Our reconstruction of stellar mass build-up is less parametric, but arguably more dependent on the underlying stellar population synthesis (SPS) models. The two results are not comparable due to the sample of galaxies studied, but the approaches are similar in terms of philosophy. An approach more similar to ours was very recently presented in \cite{EskewEtAl11}, where the authors use published estimates of the past star formation histories of three local group galaxies (obtained using resolved stellar populations) to infer where these galaxies would sit in popular diagnostic plots at previous times in their cosmic history. The work we present this paper is similar in terms of concept, but is vastly different in terms of the size and type of galaxies used, as well as overall goal.

This paper is organised as follows: in Section \ref{sec:data} we introduce the data set we use, in Section \ref{sec:method} we introduce our methodology and in Section \ref{sec:observables} we define our observables. We present our results in Section \ref{sec:results}, which we interpret and discuss in Section \ref{sec:discussion}. We summarise and conclude in Section \ref{sec:conclusions}. Throughout this paper we use assume a WMAP7 cosmology \citep{KomatsuEtAl11}.

\section{Data} \label{sec:data}

The SDSS is a photometric and spectroscopic survey, undertaken using a
dedicated 2.5m telescope in Apache Point, New Mexico. For details on
the hardware, software and data-reduction see \citet{YorkEtAl00} and
\citet{StoughtonEtAl02}. In summary, the survey was carried out on a
mosaic CCD camera \citep{GunnEtAl98}, two 3-arcsec fibre-fed
spectrographs, and an auxiliary 0.5m telescope for photometric
calibration. Photometry was taken in five bands: $u, g, r, i$ and $z$,
and Luminous Red Galaxies (LRGs) were selected for spectroscopic
follow-up according to the target algorithm described in
\citet{EisensteinEtAl01}. In this paper we analyse the latest SDSS LRG
sample (data release 7, \citealt{AbazajianEtAl09}), which includes
around 180,000 objects with a spectroscopic footprint of around 8000
sq. degrees and a redshift range $0.15 <z < 0.5$.


The target selection was designed to follow a passive stellar population in colour and apparent magnitude space, and it targets LRGs below and above $z\lesssim 0.4$ with two distinct cuts. Cut I targets low redshift LRGs by using the following cuts:

\begin{equation}
\begin{array}{l}
\displaystyle r_{p} < 13.1 + c_\parallel/0.3 \\
\displaystyle  r_{p} < 19.2 \\
\displaystyle c_\perp < 0.2 \\
\displaystyle \mu_{r,p} < 24.2 \text{ mag arcsec}^2 \\
\displaystyle r_{psf} - r_{model} > 0.3 \\
\end{array}
\end{equation}

where the two colours, $c_\parallel$ and $c_\perp$ are defined as

\begin{equation}
\begin{array}{l}
\displaystyle c_\parallel = 0.7(g-r) + 1.2[(r-i) - 0.18] \\
\displaystyle c_\perp = (r-i) - (g-r)/4 - 0.18. \\
\end{array}
\end{equation}

Model magnitudes are used for the colours, and petrosian magnitudes for the apparent magnitude and surface brightness cuts. Cut II targets LRGs at $z\gtrsim 0.4$ following:

\begin{equation}
\begin{array}{l}
\displaystyle r_p < 19.5 \\
\displaystyle c_\perp > 0.45 - (g-r)/6 \\
\displaystyle (g-r) > 1.3 + 0.25(r-i) \\
\displaystyle \mu_{r,p} < 24.2 \text{ mag arcsec}^2 \\
\displaystyle r_{psf} - r_{model} > 0.5 \\
\end{array}
\end{equation}

Two separate cuts are necessary as the passive stellar population turns sharply in a $g-r$ vs $r-i$ colour plane, when the 4000\AA\ break moves through the filters. This bend in the colour selection results in a broader colour range of targets around it.

\section{Method} \label{sec:method}

Our approach consists of tapping into the fossil record of
low-redshift ($z\lesssim0.25$) LRGs to measure their stellar evolution
in terms of their past star-formation and metallicity history out to
$z = 0.45$ (T11). As discussed in T11, using the fossil record allows us to explore
past star-formation histories in a way that is completely decoupled
from the survey's selection function. Although the SDSS selection was
designed to follow a passively evolving stellar population, it is not
true that all galaxies that pass the selection criteria follow this
model.

With the full knowledge of how the selection function of the survey
changes with redshift, we predict the number and luminosity densities
of LRGs at higher redshifts based on low-redshift samples. In the
absence of mergers, the predicted and observed number and luminosity
densities should match at high redshift. In
Section~\ref{sec:discussion} we show how, by making a few simple
assumptions, the difference between the two quantities can be
interpreted as a minimum merger history.

\subsection{Evolving galaxies back in time}

For each galaxy we consider the past history, and at which epochs it
would have been included in the survey. To do this we need to consider
the following:
\begin{enumerate}
\item the stellar evolution of the galaxy; \label{item1}
\item the selection function of the survey; \label{item2}
\item changes in the photometric errors with redshift. \label{item3}
\end{enumerate}

For~\ref{item1} we use the stellar evolution models for the LRGs of
\cite{TojeiroEtAl11}, and refer the reader to that paper for full
details of how these models were obtained. For completeness, we present a summary in Section \ref{sec:kcorrections}.

To account for \ref{item2}, we run the evolution of each individual
galaxy through the selection cuts of the LRG sample (see Section
\ref{sec:data} and \citealt{EisensteinEtAl01}), between
the redshift of the galaxy and $z=0.45$, and we construct a vector
that tells us exactly when any individual galaxy would have been
observed within the survey, were it to exist on our past light cone at
all epochs. We call this vector $V_{obs}(z)$ and we set it to unity if
the galaxy would have been observed at that redshift, and to zero
otherwise.

To incorporate the changing photometric errors (\ref{item3}) into our
colour modelling we take the following steps. First we compute the
average photometric errors in colour as a function of apparent
$r-$band cmodel magnitude, using all LRGs in the sample.  Second, for
each galaxy we compute the $g-r$ and $r-i$ offsets with respect to
their assigned model at the redshift of the galaxy - note that these
are not expected to match as each cell used to compute the 124 stacks
has a finite width in colour and redshift. We then add a term to their
predicted colour evolution which keeps the ratio between this offset
and the typical observational error at that redshift constant
throughout its whole evolution. For example, to predict the $g-r$
colour of a galaxy observed with a redshift $z_{obs}$ at any other
redshift $z > z_{obs}$ we compute
\begin{equation}\label{eq:photometric_errors}
  [g-r]_{pred}(z) = [g-r]_{model}(z) 
    + \Delta[g-r] \frac{\sigma_{g-r}[r(z)]}{\sigma_{g-r}[r(z_{obs})]},
\end{equation}
where $[g-r]_{model}$ is given by the stellar evolution models,
$\Delta[g-r]=[g-r]_{obs}-[g-r]_{model}(z_{obs})$, and $r(z)$ is the
predicted apparent $r-$band cmodel magnitude at that redshift. When
$z=z_{obs}$ this simply returns the observed colour, and the effect at
$z>z_{obs}$ is to correctly match wider areas of colour space at
larger redshift with narrower cells at low redshift. Note that if
there was no change in photometric errors with redshift there would be
no need to widen the cells.

\subsection{The stellar evolution models, K+e corrections and absolute magnitudes}\label{sec:kcorrections}

We follow \cite{TojeiroEtAl10} to compute K+e corrections and obtain
K+e corrected rest-frame absolute magnitudes - see Section 2.1 of that
paper for details - with two differences. First, we do not assume the
passive evolution model of \cite{MarastonEtAl09} but rather the
stellar evolution models of T11, which computed directly from the fossil record of LRGs. This allows us to proceed without making any assumptions about this evolution - thus making the results scientifically very robust. Secondly, a consequence of not wishing to make unsupported assumptions to construct evolutionary models is that we cannot predict how any galaxy would
look like at $z < z_{obs}$. We therefore K+e correct all galaxies to
a redshift that is larger than the high-redshift tail of our
distribution, and we choose $z_c=0.525$ (made to coincide with one of
the redshift nodes at which the k+E corrections are explicitly
computed).

There are many technical differences between the models of T11 and \cite{MarastonEtAl09} (e.g. \citealt{MarastonEtAl09} uses broadband optical colours whereas T11 uses optical spectra; the fitting philosophies are different). For the current work, the primary difference is  that T11 uses the {\em fossil record} of galaxies at a given redshift to infer their past star formation history. In contrast, \cite{MarastonEtAl09} fits a model to the colours of galaxies {\em observed over} a range of redshifts. To do so, they have to apply k+E corrections to all galaxies in order to construct a sample that is potentially coeval. The risk of such an approach is that the inferences about the evolution of the sample reflect the assumptions used to define it, entered here via the k+E corrections - they only retained galaxies at a certain redshift that they predict would have been observed at any other redshift given the survey's selection function and the assumed stellar evolution model, thus potentially considering a biased subsample of galaxies. In our approach, however, evolutionary colour paths that stray from the colour selection box {\em are} acceptable and crucially important for the results of our paper.

In T11 we assumed that that LRGs do not necessarily share the same stellar evolution or dust content, and they naturally allow for any intrinsic variation in LRGs, according to their luminosity, observed $r-i$ colour and redshift. 

Briefly, the T11 models are computed by running VESPA
\citep{TojeiroEtAl07, TojeiroEtAl09} on 124 high signal-to-noise
stacks of LRG spectra, constructed based on the observed $r-i$ colour,
luminosity and redshift of the galaxies. VESPA returns a star-formation and metallicity history, as well as a present-day dust content, for each of the 124 stacks. These stellar evolutionary
paths are turned into a spectral energy distribution, which can be
computed at any redshift larger than the redshift of the stack and
provide any set of colours, magnitudes, stellar masses or rest-frame
luminosities. 

An added subtlety lies in modelling the evolution the surface brightness of the galaxy, which is needed to fully characterise the survey selection function.  This
relies on modelling the physical size of a galaxy with redshift. Here we simply
assume the physical size of the galaxy does not change with time, and
we let the apparent size change with the angular diameter distance. In practice we know this is likely to be a simplification, as a growing body of literature discusses a potential size evolution of early-type galaxies (see the Introduction of \citealt{SaraccoEtAl11} and references within for a summary). These studies focus on the detection of a population of compact early-type galaxies at $z > 1$, with effective radii a few times smaller than the effective radii of their local counterparts of the same stellar mass (e.g. \citealt{vanDokkumEtAl08}). This does not seem to necessarily be a problem theoretically, as hydrodynamical simulations show that minor mergers can act to increase the effective radius of a galaxy by a factor of a few between $z=3$ and the present day \citep{NaabEtAl09}. Nonetheless, the difficulty in characterising the (small) samples and making the measurements at high redshift, as well as the detection of a population of compact early-type galaxies in the local Universe (e.g. \citealt{CappellariEtAl09}) makes any realistic modelling of the size of LRGs with redshift unfeasible. Given the small redshift range of our sample and, crucially, the small number of objects that are discarded solely due to failing the surface brightness cut ($< 0.01\%$), assuming no evolution in the physical size of our LRGs has a negligible effect on our results.

\section{Measured Quantities}\label{sec:observables}

The statistics that we use to test the evolution of the galaxies are
the galaxy number and luminosity densities, both as a function of
redshift. As in \cite{TojeiroEtAl10}, we use a proxy for the
luminosity of the galaxy as $L = 10^{-M_i/2.5}$, where $M_i$ is K+e
corrected to $z_c=0.525$, and we work in arbitrary units of luminosity
density throughout.

\subsection{The observed quantities}

We compute
\begin{equation}
  n_{obs}(z) = \frac{1}{V_{\Delta z}}\sum_{g \in \Delta z} \frac{1}{c_g},
\end{equation}
\begin{equation}
  \ell_{obs}(z) = \frac{1}{V_{\Delta z}}\sum_{g \in \Delta z} \frac{L_g}{c_g},
\end{equation}
where $c_g$ is an estimate of the spectroscopic completeness at the
location of each galaxy, and corrects for the small fraction of
galaxies in the target sample that were not observed \citep{PercivalEtAl07}. $\Delta z$ is the width in redshift in which $n(z)$ and $\ell(z)$ are computed. We sample  $n(z)$ and $\ell(z)$ in around 50 bins, between $z_{min} = 0.2$ and $z_{max} = 0.45$.

\subsection{The predicted quantities}

We begin by defining a redshift range, at low redshift, that we use to
predict the number and luminosity densities to higher redshifts. This
redshift bin, which we denote $\Delta z_{\rm base}$, contains galaxies
with redshifts in an interval can be varied in order to learn more
about the evolution of LRGs. For each redshift greater than those in
this bin we compute
\begin{equation}
  n_{pred}(z) = \frac{1}{V_{\Delta z_{base}}} \sum_{g\in\Delta z_{base} }
  \frac{1}{c_g}V_{obs,g}(z)
\end{equation}
\begin{equation}
  \ell_{pred}(z) = \frac{1}{V_{\Delta z_{base}}} \sum_{g\in\Delta z_{base}}
  \frac{L_g}{c_g }  V_{obs,g}(z).
\end{equation}

\subsection{Rates of change}

We define a number loss rate as
\begin{equation}\label{eq:n_loss_rate}
  r_n(z) = \frac{1}{\Delta t}\left(1 - \frac{n_{pred}(z)}{n_{obs}(z)} \right),
\end{equation}
where $\Delta t=t(\overline{\Delta z})-t(\overline{\Delta z_{base}})$, and a luminosity loss rate as
\begin{equation}\label{eq:l_loss_rate}
  r_\ell(z)  = \frac{1}{\Delta t}\left(1
    - \frac{\ell_{pred}(z)}{\ell_{obs}(z)} \right).
\end{equation}

We can compute the average luminosity lost per galaxy as
\begin{equation}\label{eq:merger_rate}
  r_g(z) =  \frac{1}{\Delta t} \left(1 
    - \frac{n_{pred}(z) / \ell_{pred}(z)} {n_{obs}(z) / \ell_{obs}(z)} \right).
\end{equation}
This is also the best estimate for the true merger rate of the
galaxies that are the progenitors of the galaxies in $\Delta
z_{base}$, assuming that any contaminants at redshift $z$ have the
same luminosity distribution as the progenitors.

\section{Results} \label{sec:results}

In this Section we show our results for a selection of $\Delta
z_{base}$ and magnitude ranges of local galaxies. If the full sample
was made of a coeval population of galaxies changing these values
would have no effect on the derived merger rates. From previous work
we know this assumption is likely to fail on two accounts: the
targeting is less robust at $z\lesssim 0.2$ as can be seen by the fact
galaxies at these redshifts tend to have small amounts of recent to
intermediate star formation and show an upturn in dust content
\citep{TojeiroEtAl11}; and the fainter objects show stronger evidence
of a non-zero merger history \citep{TojeiroEtAl10}.  As our stellar
evolution models are independent of the selection function to larger
redshift, each of these two effects should become apparent as we vary
the magnitude and the redshift of our local sample. Our interpretation
is presented in Section \ref{sec:discussion}.

\begin{figure}
\begin{center}
\includegraphics[width=3.4in]{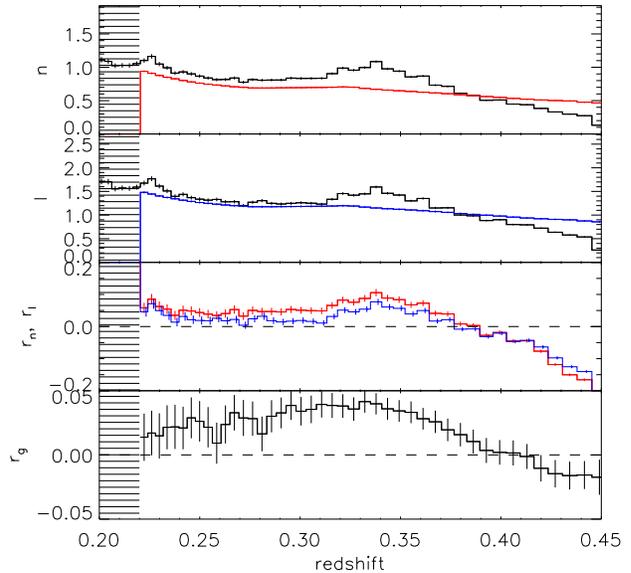}
\caption{The observed and the predicted number and luminosity
  densities (first and second panel from the top,
  respectively). Number density in units of $10^{-4}$ Mpc$^{-3} h^3$
  and the luminosity density in arbitrary units and Mpc$^{-3}
  h^3$. The observed values are in black, and the predicted values are
  in red for the number and blue for the luminosity densities. The
  shaded area shows the redshift range used to compute the predicted
  values. The third panel shows the loss rates in numbers (red) and
  luminosity (blue) given by Eqs.~\ref{eq:n_loss_rate} and
  \ref{eq:l_loss_rate}. Bottom: the galaxy luminosity loss rate given
  by Eq.~\ref{eq:merger_rate}. All rates are in Gyr$^{-1}$.}
\label{fig:rates_allM_zA}
\end{center}
\end{figure}

We begin by making no selection in luminosity, and comparing observed
and predicted densities from galaxies with $\Delta z_{base} = [0.20,
0.22]$ in the top two panels of Fig.~\ref{fig:rates_allM_zA}. On the
data side, we see a roughly constant number density from $ z \approx
0.23$ to 0.33, an excess relative to this plateau at $z \lesssim
0.23$, a bump at $z\approx0.34$, and a steep decline onwards. The
roughly constant densities are a direct result of the targeting
algorithm, which aimed to select LRGs up to a fixed number density by
following a passively evolving stellar population. The fact that these
are roughly constant can in itself put some constraints on the
dynamical evolution of the sample \citep{WakeEtAl06,WakeEtAl08}. The
excess at low redshift is likely to be contamination to the sample,
where the target algorithm has less distinguishing power. The bump at
$z \approx 0.34$ is a result of the widening colour cuts at that
redshift, as we move from cut I to cut II in the target selection. The
steep decline after that is dominated by the apparent magnitude cuts,
with the decline in the observed number and luminosity densities being
steeper than that predicted from the low redshift data.

The loss rates given by Eqs.~\ref{eq:n_loss_rate} and
\ref{eq:l_loss_rate} are shown on the third panel of
Fig.~\ref{fig:rates_allM_zA}, and we combine the information in all
four observables to compute the merger rate, or galaxy luminosity loss
rate, given by Eq.~\ref{eq:merger_rate}, which we show on the bottom
panel of the same Figure. The merger rate is roughly constant and of
the order of 3-4\%, up to the point where the sample is dominated by
the apparent magnitude limit.

\begin{figure*}
\begin{center}
\includegraphics[width=2.3in]{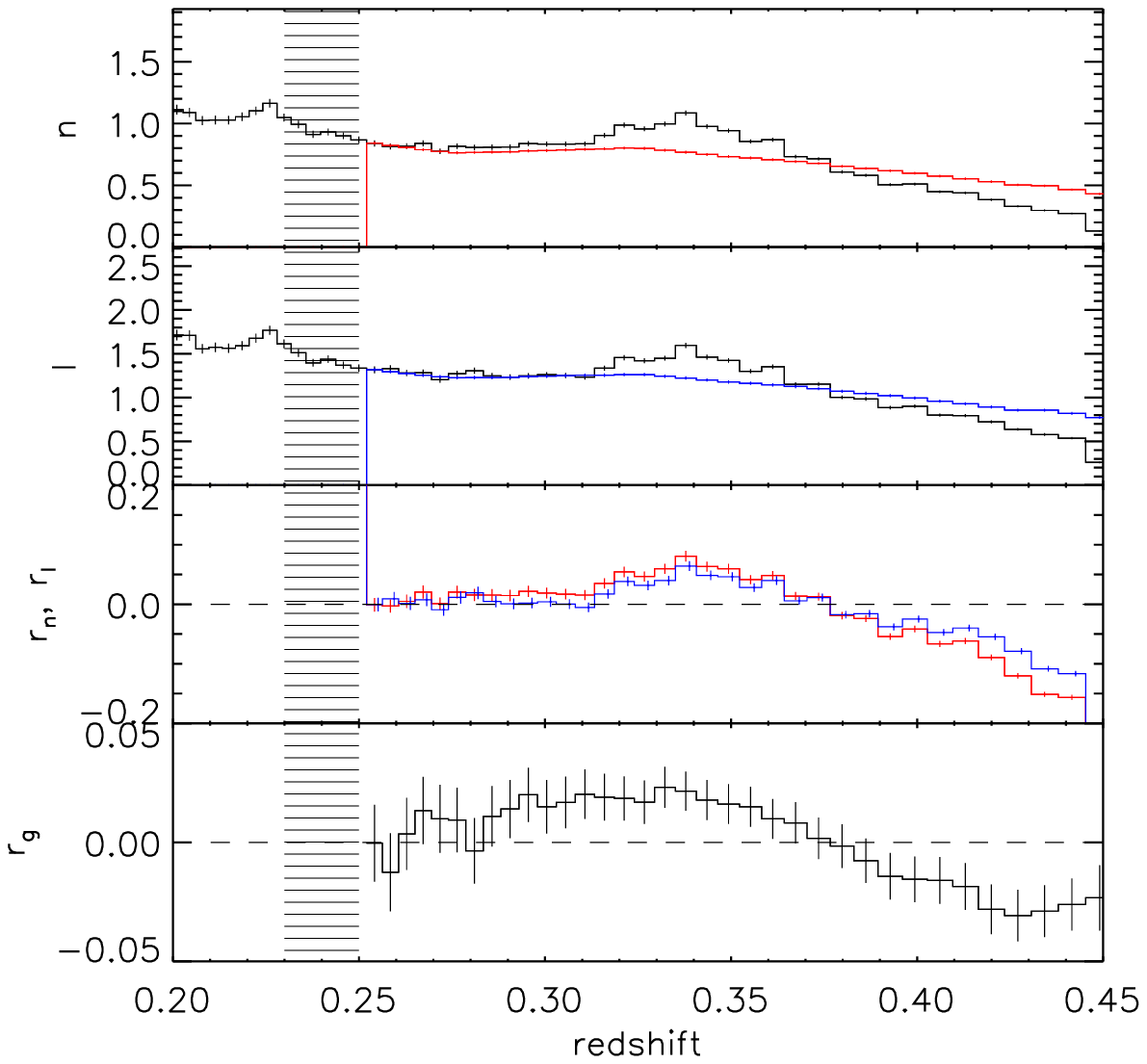}
\includegraphics[width=2.3in]{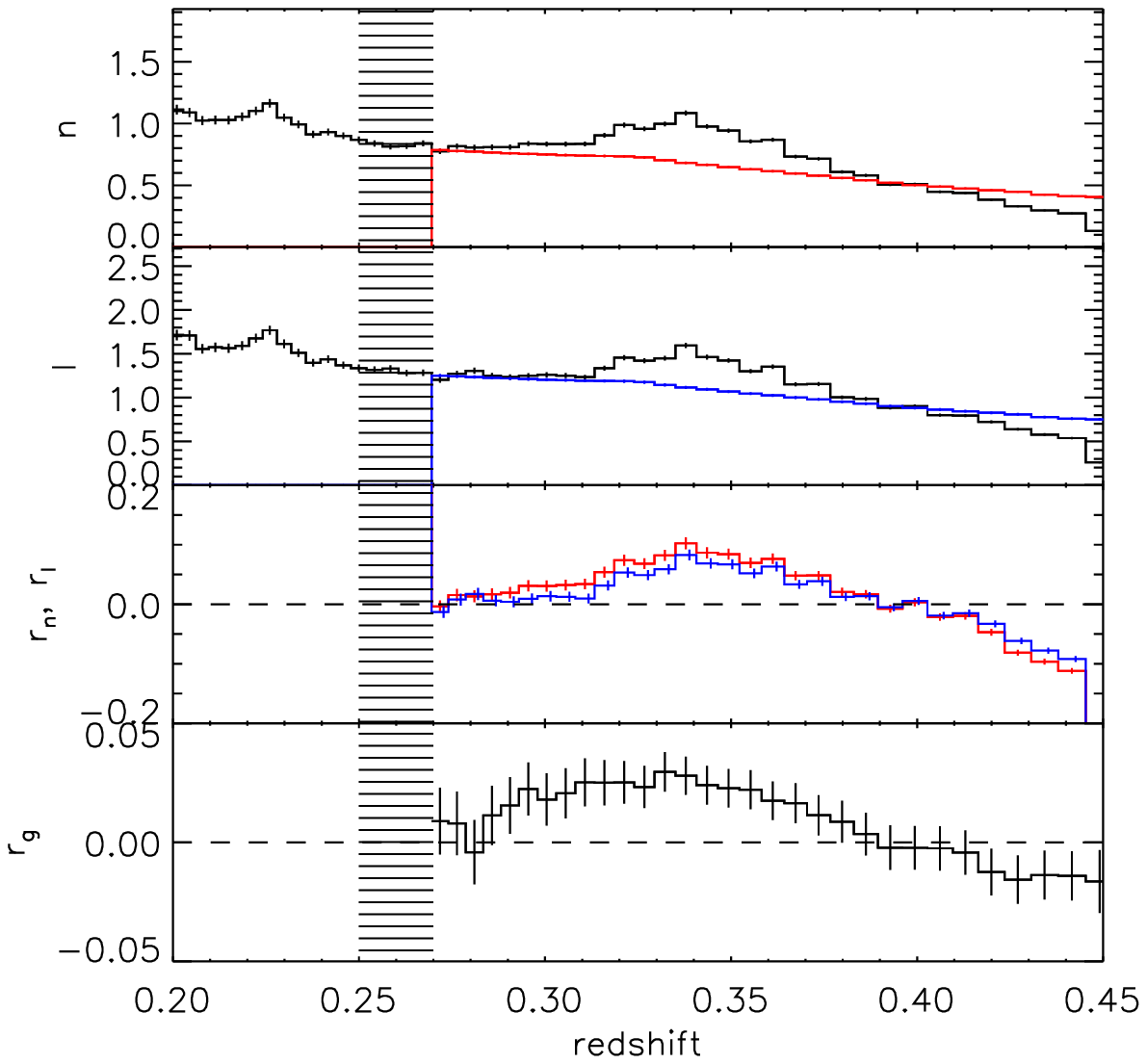}
\includegraphics[width=2.3in]{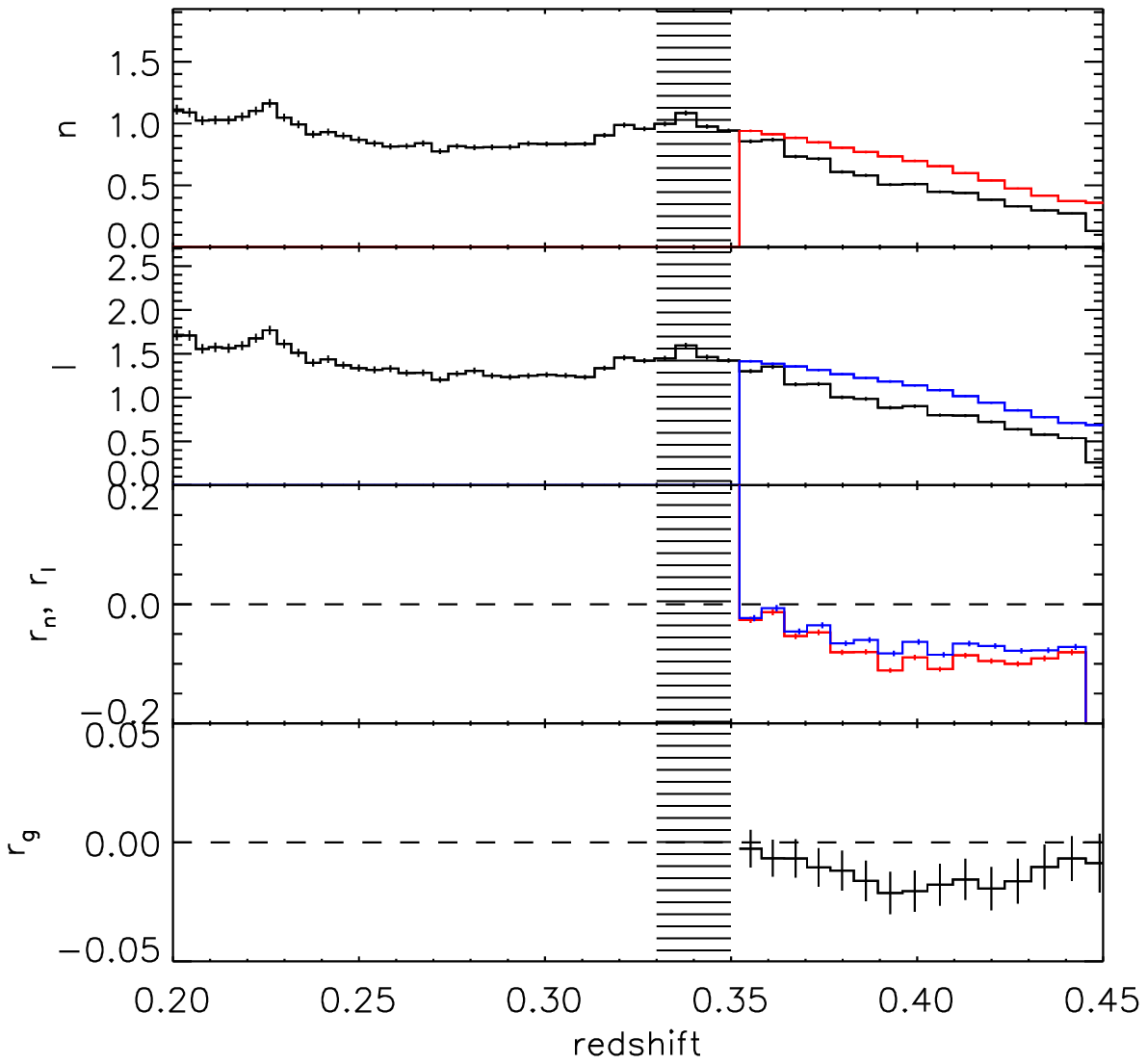}
\caption{The same as Fig.~\ref{fig:rates_allM_zA}, but exploring
  different intervals for $\Delta z_{base}$, used to compute the
  predicted quantities.}
\label{fig:rates_allM}
\end{center}
\end{figure*}

In Fig.~\ref{fig:rates_allM}, we show how the predicted quantities
change if we instead compute them from galaxies in different ranges
$\Delta z_{base}$. Note that differences in the predicted quantities
(in the blue and red lines) across the different plots provides
information about the nature and evolution of the galaxies at each of
the redshift intervals - this is discussed further in
Section~\ref{sec:discussion}.

\begin{figure*}
\begin{center}
\includegraphics[width=2.8in]{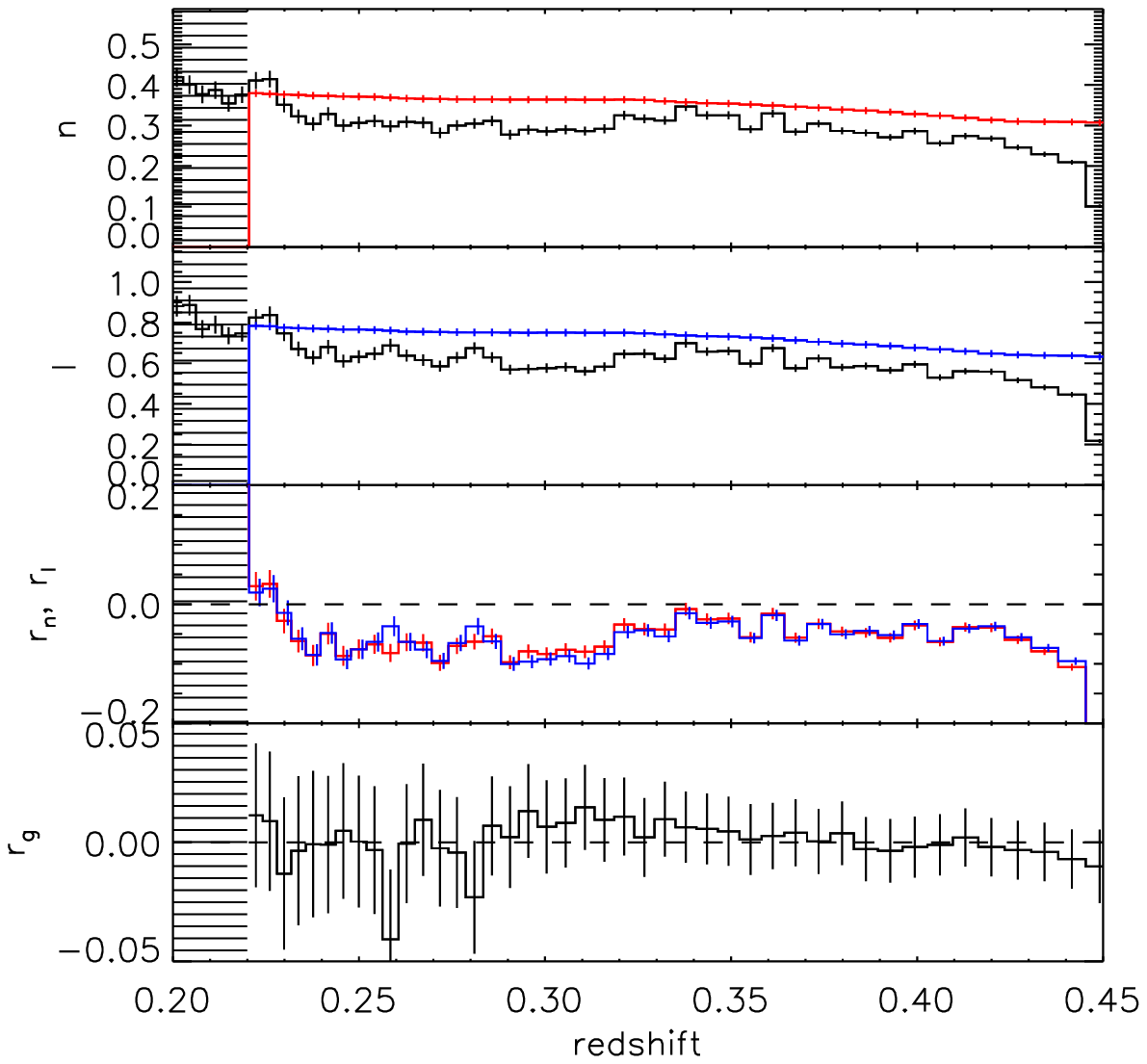}
\includegraphics[width=2.8in]{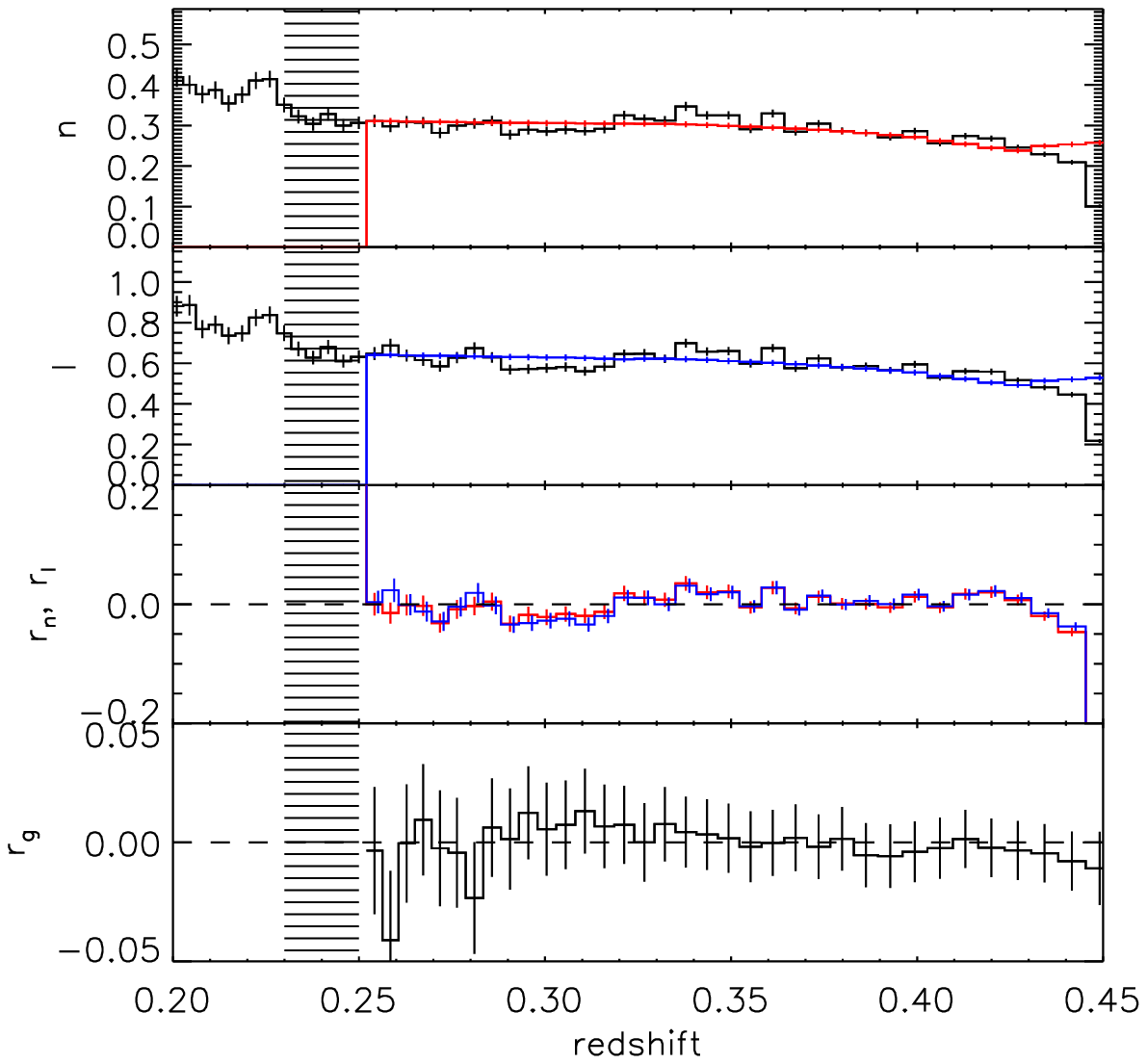}
\includegraphics[width=2.8in]{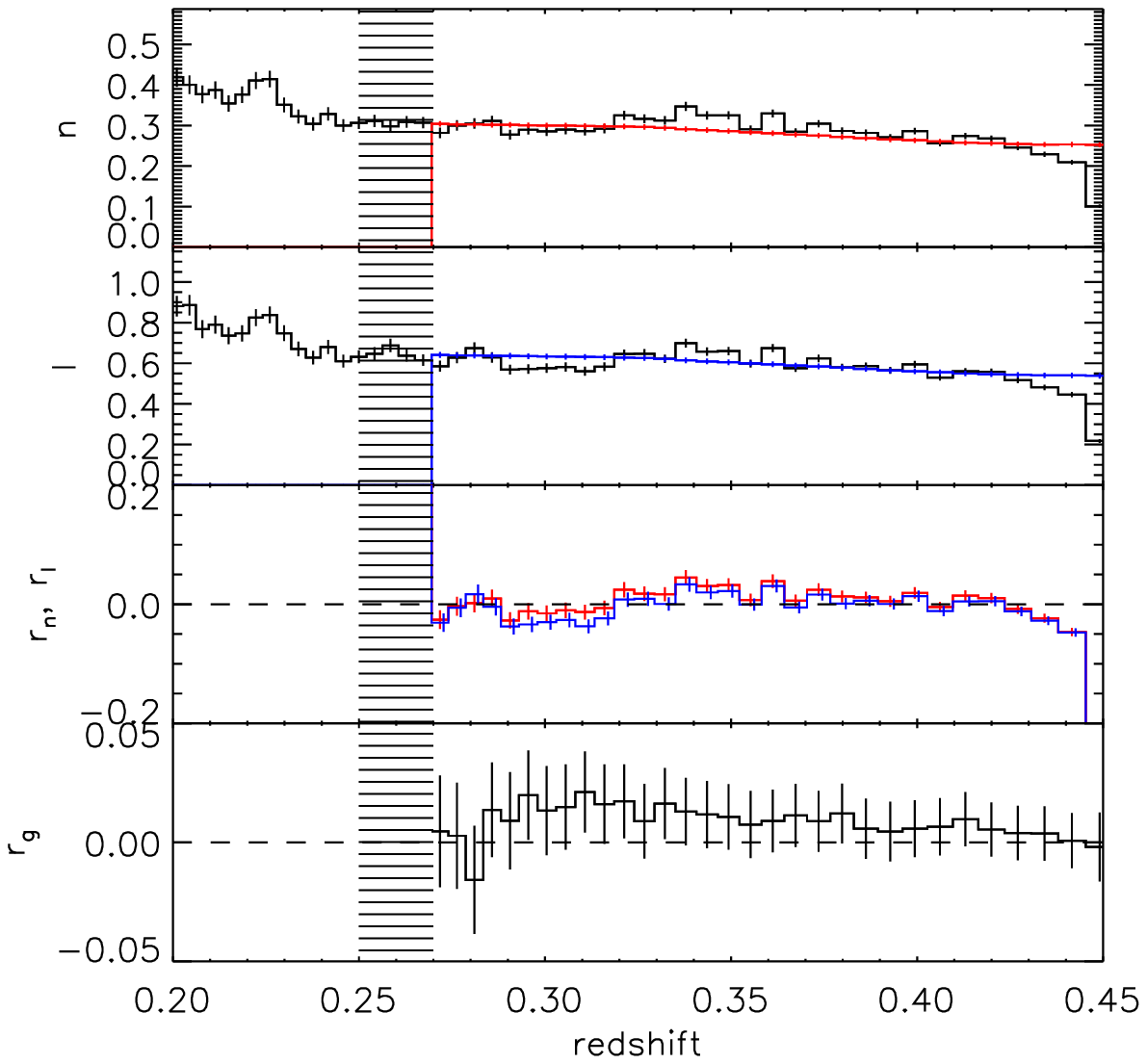}
\includegraphics[width=2.8in]{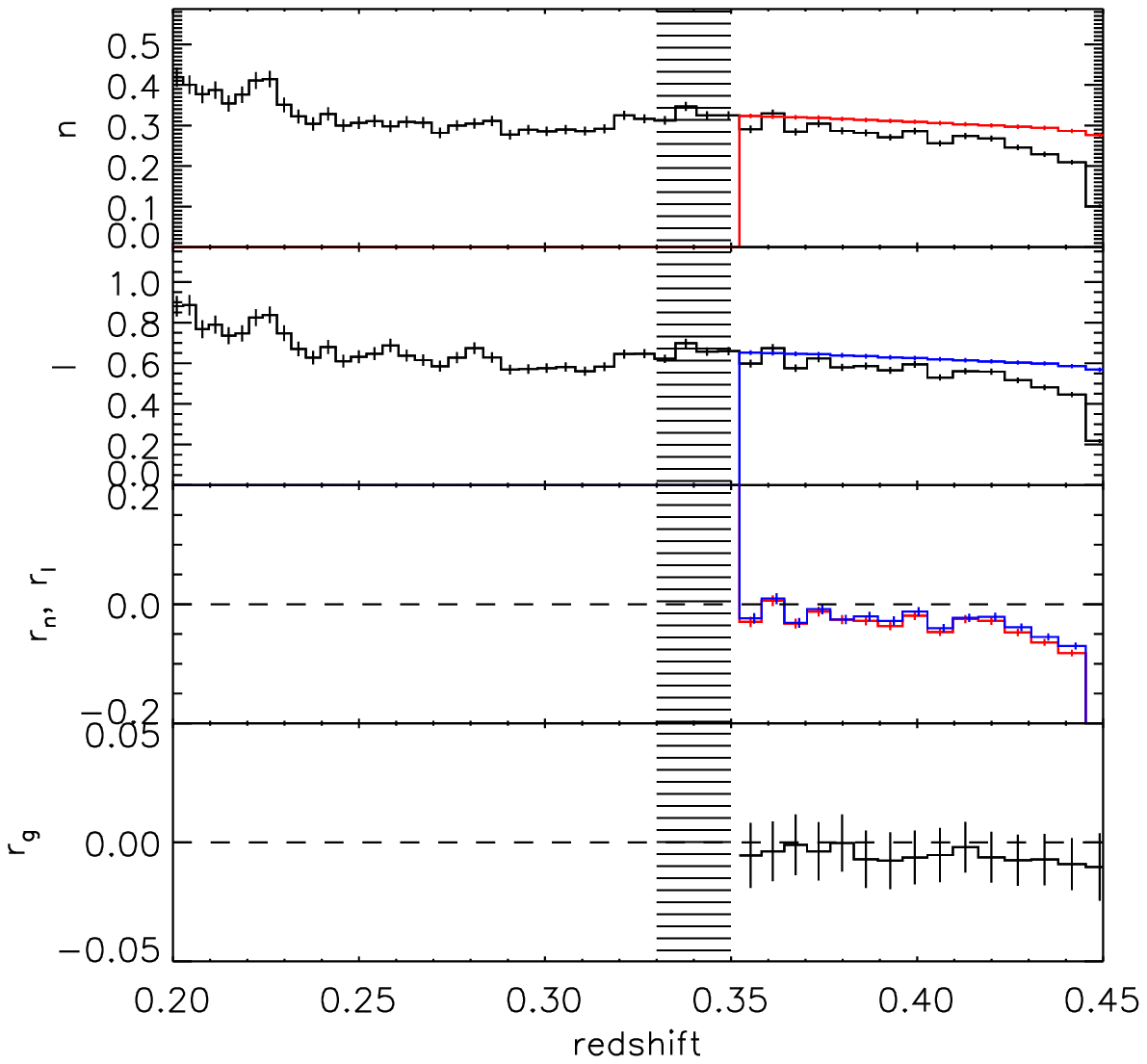}
\caption{Same as Fig.~\ref{fig:rates_allM_zA}, but using only galaxies
  with $M_i < -23$ and exploring different intervals for $\Delta
  z_{base}$, used to compute the predicted quantities.}
\label{fig:rates_M23}
\end{center}
\end{figure*}

We can also investigate how the loss rates change with the magnitude
of the galaxies. As all absolute magnitudes are K+e corrected, a flat
cut in $M_i$ with redshift selects the same population of galaxies in
the absence of mergers. In Fig.~\ref{fig:rates_M23}, we show how the
observed and predicted quantities change when we restrict the analysis
to galaxies with $M_i < -23$: this corresponds roughly to the top 30\%
brightest galaxies in the LRG sample. On the data side, we see many
features in the redshift distributions disappear, leading to a flatter
curve. The exception is at low redshift, where the excess in density
is still present. We show the same redshift ranges of $\Delta
z_{base}$ as explored in Figs.~\ref{fig:rates_allM_zA} and
\ref{fig:rates_allM}. Once again the observed curves are the same
across all four panels, with differences in the red and blue curves
holding information on the evolution of bright LRGs.  We interpret and
discuss these results in the next Section.

\subsection{Errors} \label{sec:errors}

The errors quoted throughout this paper are Poisson, and assume
that our ability to compare predicted with observed quantities depends
only how many objects we have available to do so. We account for the
change in photometric errors using Eq.~\ref{eq:photometric_errors},
which is folded into the predicted quantities. 

It is particularly hard to estimate the effect of errors from the recovered star-formation histories on the computed merger rates. In T11 we presented an extensive analysis of the errors (see their Section 3.4), which we summarise here for completeness. There are two sources of errors that can affect the recovered star formation histories from VESPA: photon noise, which perturbs the spectra in way that we assume we understand; and systematic errors in models or parametrization, which are the result of our imperfect ability to correctly model certain stages of stellar evolution, chemical enrichment or dust extinction. The former is easily estimated, via a Monte Carlo approach that we explain in \cite{TojeiroEtAl09}. In this case we are in the regime of extremely high signal-to-noise spectra and photon error is negligible. In order to understand systematic errors, in T11 we estimated these using the residuals of the best fit solution to the data (see their Fig. 3) - we found these errors to be at least 10 times larger. We use this error vector to obtain the SFHs used in the current analysis.

The effect of the uncertainties in the SFH on the measured rates comes from how many galaxies leave the survey's selection window as a function of redshift, and how bright they are. The window function of the survey is designed to follow a passive evolution, so only star-formation histories that depart significantly from this model will affect the rates. An episode of recent to intermediate star formation of anything larger than 5\% by mass is enough to take galaxies out of the survey box for up to and around 1Gyr (the exact number depends on the length of episode, the Stellar Population Synthesis (SPS) models used, the metallicity and the exact age of the burst) and, even though none of our formal error estimates allow for a discrepancy of this size, solutions obtained with different SPS models can differ by a few times this amplitude - we explicitly address the effect of using different SPS models in our results in Section~\ref{sec:SPS}.

The effect on the merger rate in turn depends on {\em how many} galaxies are removed from the sample. Our assumption in T11 is that all galaxies within a cell follow the SFH given by the stack that represents that cell. Each of our $\Delta z_{base}$ samples is typically represented by 4 to 12 stacks (depending on the luminosity range of the sample), and if one stack predicts a significant star-formation event ($\gtrsim$ 5\% by mass), then 1/12 to 1/4 of the galaxies will be predicted to leave the sample for a significant period. This means that the effect on the merger rates are somewhat quantised, rather than a smooth function of an error in the SFHs. The effect can obviously be quite dramatic, and at its extreme it can predict that almost all galaxies would leave the sample, suggesting that the observed red sequence is an intricate combination of galaxies becoming red and blue throughout cosmic history (see Section~\ref{sec:SPS} for an example).

\section{Discussion}\label{sec:discussion}

\subsection{Interpreting the observables}\label{sec:interpreting_observables}

If the sample of SDSS-II LRGs was not affected by mergers, and assuming the SFHs are correct (see Section \ref{sec:errors}), then the
observed luminosity and the number densities would obviously match the
extrapolated values. Any deviations from this, implies merger activity and/or a non-zero net flow of galaxies into the sample.
We cannot determine a balance between mergers into and
out of the sample. However, by comparing number and luminosity
evolution we can start to understand the evolution that we do see.

\renewcommand{\theenumi}{(\alph{enumi})}


\begin{figure}
\begin{center}
\includegraphics[width=4.8in]{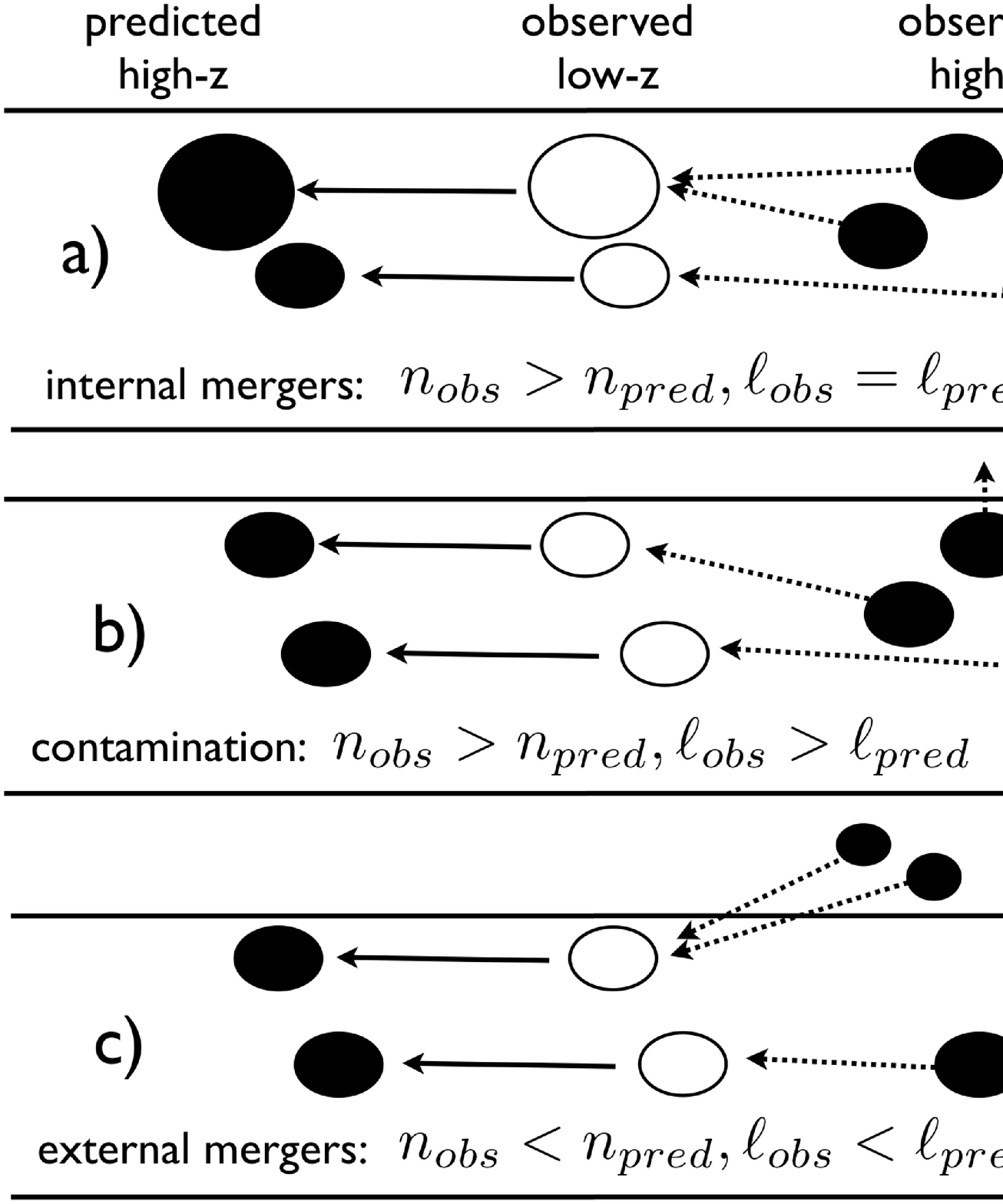}
\caption{Diagram of possible interpretations for discrepancies in the observed and predicted quantities. The middle column shows the observed luminosity and number densities. The solid arrows show the prediction given by the stellar evolution from VESPA and the survey's window function, and they lead to the left hand side column that shows the predicted luminosity and number densities. Note that whereas VESPA can only predict changes in the luminosity density, the window function can also affect the number densities. The window function is fully known and introduces no ambiguity when computing the predicted quantities - in here we show the numbers being conserved just for simplicity. The right hand side column shows the observed number and luminosity densities, and the dotted arrows represent our proposed interpretation.
See Section~\ref{sec:interpreting_observables} for more details.}
\label{fig:diagram}
\end{center}
\end{figure}

Fig.~\ref{fig:diagram} shows schematically the simplest interpretation
for three scenarios:
\begin{enumerate}
\item $n_{pred} < n_{obs}$ and $\ell_{pred} = \ell_{obs}$ - the
  simplest interpretation is that galaxies within the sample have
  merged together to decrease the predicted number density of objects,
  whilst retaining the sample luminosity. This assumes no luminosity
  loss to the intra-cluster medium.
\item shows how contamination from objects at a given redshift could
  also act to give $n_{pred} < n_{obs}$ in the same way as internal
  mergers. By contaminants here we refer to objects that would enter
  the sample for a short time only, and do not evolve to be present in
  the sample at low redshift. However, the comparison between
  $\ell_{obs}$ and $\ell_{pred}$ helps to differentiate the two
  scenarios, as in this case we expect the excess in the observed number
  density due to contamination to be matched by an excess of
  luminosity density. The two excesses will be matched in amplitude provided the contaminants have the same
  luminosity function as the galaxies that evolve through the sample.
\item $n_{pred} > n_{obs}$ and $\ell_{pred} > \ell_{obs}$. This
  scenario requires an enrichment of the sample towards lower redshift
  - either by the merging of fainter galaxies becoming brighter and
  crossing over the faint magnitude cuts, bluer galaxies turning
  redder and crossing over the colour cuts, or a combination of both.
\end{enumerate}
Note that in the case of fainter galaxies merging together to become
brighter and entering the sample by crossing over the magnitude cuts,
scenarios (a) and (c) represent the same physical process - the only
difference between is where the survey's magnitude cut lies. We
explore this fact in a little more detail in
Section~\ref{sec:mergers_LF}.

Obviously, we can expect multiple scenarios such as those in
Fig.~\ref{fig:diagram} to happen simultaneously. If this is the case, under the assumption that the luminosity function of the
contaminants matches that of the objects in the sample, and that
luminosity is conserved in a merger event, Eq.~\ref{eq:merger_rate}
gives the best estimate for the merger rate required for galaxies
within the sample between $\Delta z$ and $\Delta z_{base}$. In fact, even if
these assumptions are broken, Eq.~\ref{eq:merger_rate} still acts as a 
{\it lower limit} on the true merger rate. This is mostly easily seen from Equation (\ref{eq:merger_rate}) - we are effectively under-estimating $\ell_{pred}$, as some luminosity has been lost to the intracluster medium, resulting in an under-estimation of $r_g$.

\subsection{Interpreting the results}  \label{sec:interpret_results}

Using the results of Section~\ref{sec:results}, we can identify
redshift intervals that are likely to be dominated by each of the
scenarios shown in Fig.~\ref{fig:diagram}. Starting with the results
for all LRGS, shown in Fig.~\ref{fig:rates_allM_zA}, we see the
predicted number and luminosity densities both fall short of the
observed values up to $z\approx0.38$, suggesting a combination of
scenarios (a) and (b). The bump at $z\approx0.34$ is not reproduced by
the stellar evolution of present-day galaxies, strongly suggesting
that is it caused by contamination of galaxies as the colour cuts
widen - in other words, the wider breadth of colour in the sample at
those redshifts is not predicted by the stellar evolution of more
local galaxies. Therefore the signal here is likely dominated by
scenario (b). Finally, the sample of local galaxies predicts a much
shallower slope in the number densities as we go towards higher
redshifts, suggesting that they were still being assembled (in a
dynamical sense) at those epochs - in other words, had the local
galaxies still be in one piece at $z \gtrsim 0.4$ they would have been
bright enough to be in the sample. This puts us in a regime likely to
be dominated by scenario (c) at these high redshifts.

Moreover, at the high-redshift tail where scenario (c) dominates, we
note that the number loss rate is higher (in amplitude) than the
luminosity loss rate. In other words, the typical luminosity per
object at $z\gtrsim 0.38$ is lower than expected. This suggests that
the LRG growth due to external merging is happening predominantly at
the fainter end of the sample. We will investigate this hypothesis
further in Section \ref{sec:changing_mag}. Note that if this growth
was constant with luminosity then both loss rates would have a
negative value, but would be matched in amplitude - resulting in a
zero galaxy luminosity loss rate, interpreted in this case as a zero
internal merger rate.

\subsubsection{Changing $\Delta z_{base}$}

We now consider the effect of changing the low redshift intervals
$\Delta z_{base}$, from which we infer the galaxy evolution from high
redshift as shown in Fig.~\ref{fig:rates_allM}. Changes in the inferred
loss and merger rates as we change our low-redshift sample are
sensitive to differences in the low redshift population of galaxies.

We find no significant difference in the evolution of galaxies
observed between $z = 0.23$ and $z=0.27$ - see the first two panels of
Fig.~\ref{fig:rates_allM} - but comparing the rate of luminosity loss
per galaxy (lower panels in the figures) to those in
Fig.~\ref{fig:rates_allM_zA} suggests that the lower redshift galaxies
experience a more pronounced growth. This is consistent with a
hypothesis in which the low redshift end of the sample is not as pure
as the rest of the sample - the targeting algorithm is less robust at
these redshifts \citep{EisensteinEtAl01, TojeiroEtAl11}. Furthermore,
the sample will naturally be more enriched towards fainter galaxies,
which could have a different growth rate - see Section
\ref{sec:changing_mag} for more details.

Similarly, galaxies observed at redshifts between $0.33$ and $0.35$
are seen to have a different growth compared with galaxies at lower
redshifts. This is not unexpected, as the bump in number and
luminosity density at $z\approx0.34$ is not predicted by galaxies at
low redshifts. These results suggest that the wider colour selection
of cut II selects galaxies not only with a different stellar
evolution, but also a different merger growth rate - essentially a
different population of galaxies. For the same reasons as before, we
argue that this growth must be happening predominantly at the faint
end.

\subsubsection{Changing the luminosity}\label{sec:changing_mag}

In Fig.~\ref{fig:rates_M23} we restrict our analysis to galaxies with
a K+e corrected M$_i < -23$, which corresponds roughly to the
brightest third of the galaxies in the sample. We show results for the
same four ranges of $\Delta z_{base}$ as before. It is immediately
clear that a lot of the features in the observed quantities are less
pronounced - the bump at $z\approx0.34$ almost disappears, and the
high-redshift tail flattens out as we restrict ourselves to what is
close to magnitude-limited sample (colour cuts excluded). The
exception is the bump at low redshifts, which, combined with the
measured loss and merger rates for $\Delta z_{base} = [0.20, 0.22]$,
suggests the luminous galaxies observed at low redshift are still
growing, and this growth is approximately matches that of lower
luminosity objects. Clearly this also fits with the hypothesis that
the lower redshift galaxies form a population with more dynamical
evolution than the higher redshift galaxies within the sample.

For all other cases of $\Delta z_{base}$ we find a picture that is
consistent with passive dynamical growth, with the only - and marginal
- evidence for external growth coming for the galaxies in $\Delta
z_{base} = [0.33, 0.35]$, perhaps suggesting that not even the
brightest galaxies are purely coeval across the two LRG colour cuts
described in Section~\ref{sec:data}.

\subsection{Probing the typical luminosity of mergers}\label{sec:mergers_LF}

We noted in Section \ref{sec:interpreting_observables} that scenarios
(a) and (c) represent the same physical merger process, but can be
distinguished as (c) caused objects to cross the apparent magnitude cut
on the sample. In Section~\ref{sec:interpret_results}, we suggested
than (c) dominates at high redshift, while (a) dominates at low
redshift. We can use the information available to define a typical
luminosity for the dominant merger process happening.

\begin{figure}
\begin{center}
\includegraphics[width=3.4in]{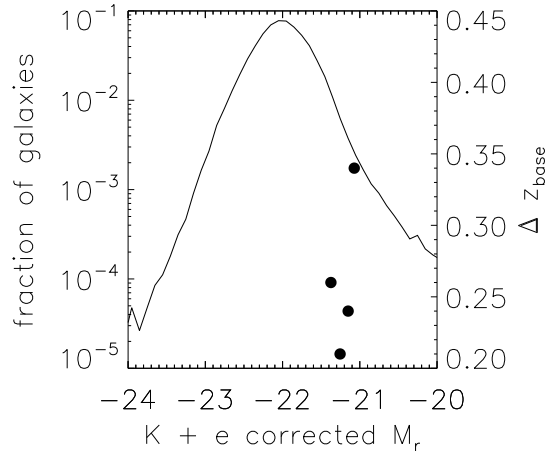}
\caption{The dots show the transitional absolute magnitude as a
  function of $\Delta z_{base}$ (on the right-hand side y-axis) - see
  Section \ref{sec:mergers_LF}. This is the absolute $r-band$
  magnitude that corresponds to the $r_p = 19.5$ cut on petrosian
  magnitude of cut II \citep{EisensteinEtAl01} at the redshifts for
  which we infer a transition from external to external growth. For
  context we also show the distribution of absolute magnitudes of
  galaxies in the sample (on the left-hand side y-axis). }
\label{fig:Mr_mergers}
\end{center}
\end{figure}

For each case of $\Delta z_{base}$ we take the redshift at which the
number and luminosity loss rates cross the zero line and compute the
absolute magnitude that would correspond to the apparent petrosian
magnitude equal to the survey's limit of $r_p = 19.5$ - in other
words, the faintest absolute magnitude observed at that
redshift. These are the galaxies that are beginning to enter the
sample when dynamical growth transitions from external to internal. Note that the two growth rates change sign at very similar epochs, so we can use either without significant effect on our results. In
Fig.~\ref{fig:Mr_mergers} we show how this transitional absolute
magnitude changes for the four different cases of $\Delta z_{base}$
(those shown in Figs.~\ref{fig:rates_allM_zA} and
\ref{fig:rates_allM}).

Once again our result is that the dynamical growth is happening
predominantly at the faint end, for galaxies observed in all the four
intervals of $\Delta z_{base}$ we studied.

\subsection{Robustness to SPS models}\label{sec:SPS}

In \cite{TojeiroEtAl11} we showed how the measured stellar evolution
of LRGs depends on the SPS models used - see Fig. 6 in that paper for
a summary of the differences. The results presented in this paper thus
far assume the stellar and colour evolution given by the Flexible
Stellar Population Synthesis (FSPS) models of \cite{ConroyEtAl09,
  ConroyAndGunn10}, which gave better fits to the spectral data and
resulted in the more passive stellar evolution model in our previous
study. However, as we could not conclusively decide which set of SPS
models gives a more accurate description of the real stellar evolution
of LRGs, in this section we discuss how the stellar evolution given by
other models would change the interpretation given above. The results
on the recovered merger rate and interpretation of the dynamical
evolution can be dramatically affected by events of star formation.

In the case of the Maraston et al. 2011 models (M10, submitted), the
amount of recent to intermediate star formation dominates only in
galaxies at $z\lesssim0.3$. The resulting interpretation is that
contamination of the LRG sample extends to a larger redshift than what
would be assumed using the FSPS models, and that these local galaxies
are not the evolutionary products of the LRGs observed at higher
redshifts. At $z>0.3$, however, the recovered dynamical evolution is
qualitatively similar to that obtained with the FSPS models, although
with slightly larger rates.

The interpretation using the stellar evolution given by the models of
\cite{BruzualEtCharlot03} (BC03), however, is incompatible with the
hypothesis that the LRGs form a coeval population of galaxies since
$z<0.45$. The significant and continuous (with redshift) amount of
recent to intermediate star formation lead to a consistently
under-prediction of the number and luminosity densities at high
redshifts from more local samples, in a way that is only explained by
a continuous amount of contamination in the sample.

\subsection{Comparison with previous results}

The results in this paper are in qualitative agreement with those in
\cite{TojeiroEtAl10}, where we used the luminosity weighted
power-spectrum to argue that the LRG population is growing via
external mergers at the faint end. Quantitatively, however, the rates
we find in this paper are lower by roughly a factor of two. The main
difference between the two approaches lies on the treatment of the
stellar evolution - whilst in \cite{TojeiroEtAl10} we assumed that the
passive stellar evolution model of \cite{MarastonEtAl09} was suitable
for all galaxies in the sample, in here we instead use the stellar
evolution models of \cite{TojeiroEtAl11} which use the fossil record
of LRGs of different colour, luminosity and redshift to infer the
most-likely stellar evolution. Furthermore, in \cite{TojeiroEtAl10},
we assumed that the scatter in colour and magnitude was due to
intrinsic or photometric errors, which is a byproduct of assuming that
all LRGs in the sample share the same stellar evolution. Instead, the
analysis presented in this paper makes no such assumption, and is
robust to an influx and/or an outflux of objects if we interpret the
measured rates as a lower limit. It is the independent treatment of
the stellar evolution, and the detailed comparison of predicted with
observed quantities that allows us to make an assessment on the nature
of the mergers without the need for a clustering analysis.

Comparisons with other works is less trivial, due to the typically
different samples used. We refer the reader to Section 8 of
\cite{TojeiroEtAl10} for a summary of recent results and
comparison. We continue to argue that merging at the faint end of the
LRG population is a more likely explanation published for the
published evidence for both luminosity growth \citep{BrownEtAl07,
  WhiteEtAl07, CoolEtAl08, MasjediEtAl08} and loss of number density
\citep{MasjediEtAl06, WakeEtAl06, WhiteEtAl07, WakeEtAl08,
  deProprisEtAl10}.

\section{Summary and conclusions} \label{sec:conclusions}

In this paper we have demonstrated how we can use the fossil record of
galaxies to extrapolate from low redshift samples to match samples
selected at higher redshift. By comparing against the observed number
and luminosities, we can measure growth and consider where the
dynamical evolution of galaxies is required. We have applied this
technique to consider the dynamical evolution of galaxies in the
SDSS-II LRG sample, although the technique could be used to link
different surveys at low and high redshift.

For the SDSS-II LRG sample, we have previously predicted the stellar
evolution in \cite{TojeiroEtAl11}, by using the fossil record encoded
in the spectra of LRGs. This allowed us to compute the most likely
stellar evolution of observed LRGs in a way that was decoupled from
the survey's selection function. We used this information to make
predictions on the number and luminosity of LRGs at higher redshift
from a local sample of galaxies, and we showed how differences in
these quantities can be interpreted as a minimum merger rate.
Our main conclusions can be summarised as the two following points:
\begin{itemize}
\item the LRG population is not purely coeval, with some of galaxies
  targeted at $z<0.22 $ and at $z>0.34$ showing different dynamical
  growth than galaxies targeted throughout the sample;
\item the LRG population is still dynamically growing, and this growth
  must be limited to the faint end;
\end{itemize}

Based on these results, we argue that a coeval population of galaxies
that has been dynamically passive to less than 1-2\% can be robustly
selected from the LRG sample, by imposing a cut on the K+e corrected
absolute magnitude, and by selecting galaxies with $z \in [0.23,
0.45]$. This merger rate is lower than those previously measured by roughly a factor of two,
because, by using empirically determined evolutionary tracks we no
longer need to assume passive evolution, and this previously led to incorrect
evolutionary tracks for some galaxies.

We caution against interpreting these results beyond the limits of the stellar population models, and once again stress the important of good models and statistically good fits for analyses of this type. However, given the power inherent in the type of approach, and improvements in SPS modelling due in the future, we consider this analysis to be a forerunner of significant future work.

\section{Acknowledgments}

We thank the anonymous referee for a report that helped us clarify a number of issues. We thank David Wake for helpful discussions and encouragement. RT thanks Alan Heavens and Raul Jimenez for help in the development of VESPA.
RT thanks the Leverhulme trust for financial support. WJP is
grateful for support from the UK Science and Technology Facilities
Council (grant ST/I001204/1), the Leverhulme trust and the European Research Council. 

Funding for the SDSS and SDSS-II has been provided by the Alfred
P. Sloan Foundation, the Participating Institutions, the National
Science Foundation, the U.S. Department of Energy, the National
Aeronautics and Space Administration, the Japanese Monbukagakusho, the
Max Planck Society, and the Higher Education Funding Council for
England. The SDSS Web Site is http://www.sdss.org/.

\bibliographystyle{mn2e}
\bibliography{/Users/ritat//WORK/LRG/PAPER/my_bibliography}

\end{document}